\documentclass[12pt]{iopart}

\usepackage{iopams}  
\usepackage{amssymb}
\eqnobysec

\begin{document}

\title[]{The generalized Heun equation in QFT in curved space-times}

\author{D. Batic}
\address{Institute for Theoretical Physics, ETH Z\"{u}rich, CH-8093 Z\"{u}rich, Switzerland}
\ead{batic@itp.phys.ethz.ch}
\author{H. Schmid}
\address{UBH Software \& Engineering GmbH, D-92224 Amberg, Germany}
\ead{Harald.Schmid@UBH.de}
\author{M. Winklmeier}
\address{FB 3-Mathematik, Universit\"{a}t Bremen, D-28359 Bremen, Germany}
\ead{winklmeier@math.uni-bremen.de}

\begin{abstract}
In this article we give a brief outline of the applications of the generalized Heun equation (GHE) in the context of Quantum Field Theory in curved space-times. In particular, we relate the separated radial part of a massive Dirac equation in the Kerr-Newman metric and the static perturbations for the non-extremal Reissner-Nordstr\"{o}m solution to a GHE. 

\end{abstract}

\submitto{\JPA}
\maketitle

\section{Introduction}\label{sec:1}
According to Sch\"{a}fke and Schmidt \cite{sch} the generalized Heun equation (GHE) is a second order differential equation of the form
\begin{equation}\label{GHE}
\fl
y^{''}(z)+\left(\frac{1-\mu_{0}}{z}+\frac{1-\mu_{1}}{z-1}+\frac{1-\mu_{2}}{z-\widehat{a}}+\alpha\right)y^{'}(z)+\frac{\beta_{0}+\beta_{1}z+\beta_{2}z^{2}}{z(z-1)(z-\widehat{a})}y(z)=0
\end{equation}
where $\widehat{a}\in\mathbb{C}\backslash\{0,1\}$ and $\mu_{0}$, $\mu_{1}$, $\mu_{2}$, $\alpha$, $\beta_{0}$, $\beta_{1}$, $\beta_{2}$ are arbitrary complex numbers. Moreover, $0$, $1$ and $\widehat{a}$ are simple singularities with exponents $\{0,\mu_{0}\}$, $\{0,\mu_{1}\}$ and $\{0,\mu_{2}\}$ respectively, while $\infty$ is at most an irregular singularity of rank $1$. Following \cite{sch} we choose parameters $\Lambda:=(\Lambda_{1},\Lambda_{2},\Lambda_{3})\in\mathbb{C}^{3}$ such that the last coefficient of (\ref{GHE}) takes the form
\[
\fl
\frac{\beta_{0}+\beta_{1}z+\beta_{2}z^{2}}{z(z-1)(z-\widehat{a})}=\sum_{\sigma,\rho=0 \atop \sigma\neq\rho}^2\frac{1}{2}\left(\frac{1-\mu_{\sigma}}{z-z_{\sigma}}\right)\left(\frac{1-\mu_{\rho}}{z-z_{\rho}}\right)+\sum_{k=0}^{2}\frac{(\alpha/2)(1-\mu_{k})+\Lambda_{k}}{z-z_{k}}
\]
with $z_{0}:=0$, $z_{1}:=1$, $z_{2}:=\widehat{a}\in\mathbb{C}\backslash\{0,1\}$. Note that for fixed $\mu_{0}$, $\mu_{1}$, $\mu_{2}$, $\alpha$ and $\widehat{a}$ $\Lambda$ is uniquely determined by $(\beta_{0},\beta_{1},\beta_{2})$ and vice versa.\\
To underline the importance of equation (\ref{GHE}) we recall that it contains the ellipsoidal wave equation as well as the Heun equation ($\alpha=\beta_{2}=0$) and thus the Mathieu, spheroidal, Lam$\acute{\mbox{e}}$, Whittaker-Hill and Ince equations as special cases. The aim of our work is to present some specific examples where the GHE arises in physics with particular attention to QFT theory in curved space-times.\\
Sch\"{a}fke and Schmidt obtained under certain conditions on $\widehat{a}$ all connection coefficients between the Floquet solutions at the singularities $0$, $1$ and $\widehat{a}$, thus determining the full monodromy group of (\ref{GHE}). Furthermore, they showed that the connection coefficients can be computed by four-term recurrence relations.\\
Notice that for $\alpha=0$ and $\beta_{2}=\Lambda_{1}+\Lambda_{2}+\Lambda_{3}=0$ the equation (\ref{GHE}) becomes the Heun equation where the point at infinity is a simple singularity with exponents
\[
\{\nu+1-(1/2)(\mu_{0}+\mu_{1}+\mu_{2}),-\nu+1-(1/2)(\mu_{0}+\mu_{1}+\mu_{2})\}
\]
and
\[
\nu^{2}=\frac{1}{4}\left(\mu_{0}^{2}+\mu_{1}^{2}+\mu_{2}^{2}\right)-\frac{1}{2}-\Lambda_{1}-\widehat{a}\Lambda_{2}.
\]

\section{The Dirac equation in the Kerr-Newman metric and the GHE}\label{sec:2}
In Boyer-Lindquist coordinates $(t,r,\vartheta,\varphi)$ with $r>0$, $0\leq\vartheta\leq\pi$, $0\leq\varphi<2\pi$ the Kerr-Newman metric is given by \cite{wald}
\[
\fl
\eqalign{
\rmd s^{2}\ =\ &
\frac{\Delta-a^{2}\sin^{2}{\vartheta}}{\Sigma}\ \rmd t^{2}
+\frac{2a\sin^{2}{\vartheta}(r^2+a^2-\Delta)}{\Sigma}\ \rmd t\, \rmd \varphi
-(r^2+a^2)^{2}\,\sin^2{\vartheta} \frac{\widetilde{\Sigma}}{\Sigma}\ \rmd \varphi^{2}\\
&-\frac{\Sigma}{\Delta}\ \rmd r^2
-\Sigma\ \rmd \vartheta^{2}
}
\]
with
\[
\Sigma:=\Sigma(r,\theta)=r^2+a^2\cos^{2}\theta, \qquad \Delta:=\Delta(r)=r^2-2Mr+a^2+Q^2
\]
and
\[
\widetilde{\Sigma}:=\widetilde{\Sigma}(r,\vartheta)=1-a^2\gamma^{2}(r)\sin^{2}{\vartheta},\qquad \gamma(r):=\frac{\sqrt{\Delta}}{r^2+a^2}
\]
where $M$, $a$ and $Q$ are the mass, the angular momentum per unit mass and the charge of the black hole, respectively. In the non-extreme case $M^2>a^2+Q^2$ the function $\Delta$ has two distinct zeros, namely,
\[ 
r_{-}=M-\sqrt{M^2-a^2-Q^2}, \qquad r_{+}=M+\sqrt{M^2-a^2-Q^2},
\]
the first one corresponding to the Cauchy horizon and the second one to the event horizon. Moreover, $\Delta>0$ for $r>r_{+}$. In the extreme case $M^2=a^2+Q^2$ the Cauchy horizon and the event horizon coincide since $\Delta$ has a double root at $r_{+}^{*}=M$.
According to Penrose and Rindler \cite{pen} the Dirac equation coupled to a general gravitational field $\textbf{V}$ is given in terms of two-component spinors $(\phi^{A},\chi^{A^{'}})$ by 
\[
(\nabla^{A}_{A^{'}}-\rmi eV^{A}_{A^{'}})\phi_{A}=\frac{m_{e}}{\sqrt{2}}\chi_{A^{'}},\qquad   
(\nabla_{A}^{A^{'}}-\rmi eV_{A}^{A^{'}})\chi_{A^{'}}=\frac{m_{e}}{\sqrt{2}}\phi_{A}
\]
where we used Planck units $\hbar=c=G=1$. Furthermore, $\nabla_{AA^{'}}$ is the symbol for covariant differentiation, $e$ is the charge or coupling constant of the Dirac particle to the vector field $\textbf{V}$ and $m_{e}$ is the particle mass.
The Dirac equation in the Kerr-Newman geometry was computed and separated by Page \cite{page} with the help of the Kinnersley tetrad \cite{kin}. In view of the separation of the Dirac equation we choose to work with the Carter tetrad \cite{carter}. If we make the ansatz \cite{chandra,KN}
\[
\Psi=\left( \begin{array}{c}
               -\phi_{0}\\
               +\phi_{1}\\
               \rmi\chi_{1^{'}}\\
               \rmi\chi_{0^{'}}
\end{array}\right)=\rme^{\rmi\omega t}\rme^{\rmi\widehat{k}\varphi}S(r,\vartheta)\left( \begin{array}{c}
               R_{-}(r)S_{-}(\vartheta)\\
               R_{+}(r)S_{+}(\vartheta)\\
               R_{+}(r)S_{-}(\vartheta)\\
               R_{-}(r)S_{+}(\vartheta)
\end{array} \right),\quad\widehat{k}=k+\frac{1}{2},
\]
where $\omega$ and $k\in\mathbb{Z}$ denote the energy and the azimuthal quantum number of the particle, respectively, and $S$ is the non singular matrix 
\[
\fl
S(r,\vartheta)=\Delta^{-\frac{1}{4}}~\mbox{\upshape{diag}}\left(\frac{1}{\sqrt{r-\rmi a\cos{\vartheta}}},\frac{1}{\sqrt{r-\rmi a\cos{\vartheta}}}, \frac{1}{\sqrt{r+\rmi a\cos{\vartheta}}},\frac{1}{\sqrt{r+\rmi a\cos{\vartheta}}}\right),
\]
then the Dirac equation decouples into the following systems of linear first order differential equations for the radial $R_{\pm}$ and angular components $S_{\pm}$ of the spinor $\Psi$ 
\begin{eqnarray} 
&&\left( \begin{array}{cc}
     \sqrt{\Delta}\widehat{\mathcal{D}}_{-}&-\rmi m_{e}r-\lambda\\
     \rmi m_{e}r-\lambda&\sqrt{\Delta}\widehat{\mathcal{D}}_{+}
           \end{array} \right)\left( \begin{array}{cc}
                                     R_{-} \\
                                     R_{+}
                                     \end{array}\right)=0, \label{radial}\\
&&\left( \begin{array}{cc}
     -\widehat{\mathcal{L}}_{-} & \lambda+am_{e}\cos\theta\\
                \lambda-am_{e}\cos\theta & \widehat{\mathcal{L}}_{+}
           \end{array} \right)\left( \begin{array}{cc}
                                     S_{-} \\
                                     S_{+}
                                     \end{array}\right)=0 \label{angular}
\end{eqnarray} 
where
\begin{eqnarray}
\fl
&&\widehat{\mathcal{D}}_{\pm}=\frac{\rmd}{\rmd r}\mp\rmi\frac{K(r)}{\Delta},\quad \hspace{1.5cm}K(r)=\omega(r^2+a^2)-eQr+a\widehat{k}, \label{ramb1}\\
\fl
&&\widehat{\mathcal{L}}_{\pm}=\frac{\rmd}{\rmd\vartheta}+\frac{1}{2}\cot{\vartheta}\pm Q(\vartheta),\quad Q(\vartheta)=a\omega\sin{\vartheta}+\widehat{k}\csc{\vartheta} \label{ramb2}.
\end{eqnarray}
For a detailed derivation of (\ref{radial}) and (\ref{angular}) we refer to \cite{KN}. Let us briefly recall that in the case $a=0$ the components $S_{\pm}$ of the angular eigenfunctions can be expressed in terms of spin-weighted spherical harmonics \cite{pen1,gold} whereas for $a\neq 0$ they satisfy a generalized Heun equation \cite{uns}. We show now how to transform the radial equations for the components $R_{\pm}$ of the radial functions in such a way that they are reduced to a GHE. In what follows a prime $^{'}$ always denotes differentiation with respect to $r$. First of all let us bring (\ref{ramb1}) into the form
\begin{eqnarray}
R^{'}_{-}(r)+\nu(r)R_{-}(r)&=&f(r)R_{+}(r) \label{uno}\\
R^{'}_{+}(r)+\overline{\nu}(r)R_{+}(r)&=&\overline{f}(r)R_{-}(r) \label{due}
\end{eqnarray}
with
\[
\nu(r):=\rmi\frac{K(r)}{\Delta(r)},\quad f(r):=\frac{\lambda+\rmi m_{e}r}{\sqrt{\Delta(r)}}.
\]
From the above system and the definitions for $\nu$ and $f$ it can be immediately verified that $\overline{R}_{\pm}=R_{\mp}$. As a consequence we just need to derive the ODE satisfied by $R_{-}$. A simple computation gives
\[
R^{''}_{-}+pR^{'}_{-}+qR_{-}=0,\quad R^{''}_{+}+\overline{p}R^{'}_{+}+\overline{q}R_{+}=0
\]
with
\[
p:=-\frac{f^{'}}{f},\quad q:=|\nu|^{2}-|f|^{2}+p\nu+\nu^{'}
\]
where to keep notation simple we write $R_{\pm}$, $p$, $q$, $f$, $\nu$ instead of $R_{\pm}(r)$ and so on. Notice that $q\to\omega^{2}-m^{2}_{e}$ for $r\to+\infty$ as can be seen easily from the expansion in partial fractions below. Moreover, we shall restrict our attention to the case $|\omega|>m_{e}$. For future convenience we write $p$ and $q$ in terms of partial fractions as follows
\[
p(r)=\frac{p_{0}}{r-r_{0}}+\frac{p_{-}}{r-r_{-}}+\frac{p_{+}}{r-r_{+}}
\]
where
\[
p_{0}=-1,\quad p_{-}=p_{+}=\frac{1}{2},\quad r_{0}=\rmi\frac{\lambda}{m_{e}}
\] 
and 
\begin{eqnarray*}
   q(r)\, &=\,
   T + \frac{T_0}{r-r_0} + \frac{T_-}{r-r_-} + \frac{T_+}{r-r_+}
   + \frac{S_0}{(r-r_0)^2} + \frac{S_-}{(r-r_-)^2} + \frac{S_+}{(r-r_+)^2}.
\end{eqnarray*}
where

\begin{eqnarray*}
   T\, &=\, \omega^2 -m_e^2,
   \\[1ex]
   T_0\, &=\, -\rmi \omega^2 
   + \frac{\rmi A}{r_+-r_0} + \frac{\rmi B}{r_+-r_0},
   \\[1ex]
   T_-\, &=\, 2B\omega^2 
   - \frac{2AB -m_e^2 r_-^2 - \lambda^2}{r_+-r_-}
   + \frac{\rmi\omega}{2}
   - \frac{\rmi(A+B)}{2(r_+-r_-)}
   - \frac{\rmi B}{r_--r_0},
   \\[1ex]
   T_+\, &=\, 2A\omega^2 
   - \frac{2AB -m_e^2 r_+^2 - \lambda^2}{r_+-r_-}
   + \frac{\rmi\omega}{2}
   + \frac{\rmi(A+B)}{2(r_+-r_-)}
   - \frac{\rmi A}{r_+-r_0},
   \\[3ex]
   S_0\, &=\, 0,
   \qquad
   S_-\, =\, B^2 - \frac{\rmi B}{2},
   \qquad
   S_+\, =\, A^2 - \frac{\rmi A}{2}
\end{eqnarray*}
with
\begin{eqnarray*}
   A\, &=\, \frac{K(r_+)}{r_+-r_-},
   \qquad
   B\, &=\, -\frac{K(r_-)}{r_+-r_-}.
\end{eqnarray*}

 Finally, by transforming $R_{-}$ according to
\[
R_{-}(r)=(r-r_{-})^{\iota_{-}}(r-r_{+})^{\iota_{+}}e^{\widehat{\gamma}r}\widehat{R}_{-}(r)
\]
and introducing the new variable $z\in\mathbb{C}$ defined by
\[
z=\frac{r-r_{-}}{r_{+}-r_{-}},
\]
it can be checked that if $\iota_{\pm}$ and $\widehat{\gamma}$ are such that
\begin{eqnarray*}
&&\widehat{\gamma}=\pm\rmi\sqrt{\omega^{2}-m^{2}_{e}},\\
&&\iota^{2}_{\pm}+(p_{\pm}-1)\iota_{\pm}+S_{\pm}=0,
\end{eqnarray*}
then (\ref{uno}) reduces to (\ref{GHE}) with
\begin{eqnarray*}
&&\mu_{0}=\frac{1}{2}-2\iota_{-},\quad \mu_{1}=\frac{1}{2}-2\iota_{+},\quad \mu_{2}=2,\quad\alpha=2(r_{+}-r_{-})\widehat{\gamma},\\
&&\Lambda_{0}=(r_{+}-r_{-})T_{-},\quad\Lambda_{1}=(r_{+}-r_{-})T_{+},\quad\Lambda_{2}=(r_{+}-r_{-})T_{0}\\
&&\widehat{a}=-\frac{1}{2}+\frac{M}{r_{+}-r_{-}}+\rmi\frac{\lambda/m_{e}}{r_{+}-r_{-}}.
\end{eqnarray*}
  
\section{Static perturbations of the non extremal Reissner-Nordstr\"{o}m solution and the GHE}
\label{sec:3}
Let us consider the following Lagrangian describing four-dimensional gravity coupled to a $U(1)$ gauge field and a real massive scalar, with a non-renormalizable but gauge invariant coupling between the gauge field and the scalar, namely
\begin{eqnarray*}
&&g^{-1/2}\mathcal{L}=\frac{R}{16\pi G_{N}}-\frac{1}{2}(\partial_{\mu}\phi)^{2}-\frac{f(\phi)}{4}F^{2}_{\mu\nu}-V(\phi),\\
&&V(\phi)=\frac{1}{2}m^{2}\phi^{2},\quad f(\phi)=\frac{1}{1+\ell^{2}\phi^{2}}
\end{eqnarray*}
where $R$ and $G_{N}$ are the scalar curvature and the Newton constant, respectively, and $\ell$ is a parameter having the dimensions of length. In order to construct static solutions with magnetic charge $g$ we require that
\begin{eqnarray*}
&&ds^{2}=g_{tt}dt^{2}+g_{rr}dr^{2}+r^{2}(d\theta^{2}+\sin^{2}\theta~d\varphi^{2}),\\
&&F_{2}=\frac{1}{2}F_{\mu\nu}~dx^{\mu}\wedge dx^{\nu}=4\pi g=g~d\theta\wedge \sin{\theta}~d\varphi,\\
&&\phi=\phi(r).
\end{eqnarray*}
Moreover, 
\begin{eqnarray*}
&&\square\phi=\frac{1}{\sqrt{g}}\partial_{r}\sqrt{g}g^{rr}\partial_{r}\phi=\frac{\partial V_{eff}}{\partial\phi},\quad V_{eff}(\phi,r)=V(\phi)+\frac{g^{2}}{2r^{4}}f(\phi),\\
&&G_{\mu\nu}=8\pi G_{N}T_{\mu\nu}.
\end{eqnarray*}
By setting $M_{Pl}\equiv 1/\sqrt{8\pi G_{N}}=1$ and
\[
g_{tt}=-\rme^{2A(r)},\quad g_{rr}=\rme^{2B(r)}
\]
we obtain for $\phi$ and $B$ the following system of equations \cite{gup}
\begin{eqnarray*}
&&\phi^{''}+\left(\frac{2}{r}-2B^{'}+\frac{1}{2}r{\phi^{'}}^{2}\right)=\rme^{2B}\frac{\partial V_{eff}}{\partial\phi},\\
&&\frac{1}{2}{\phi^{'}}^{2}+\rme^{2B}V_{eff}-\frac{2B^{'}}{r}+\frac{1-\rme^{2B}}{r^{2}}=0.
\end{eqnarray*} 
Linearizing the above equations of motion around $\phi=0$ and requiring that the horizon is at $r_{H}=1$ yields
\begin{eqnarray*}
&&\phi^{''}+\left(\frac{2}{r}-2B^{'}\right)\phi^{'}=\rme^{2B}\frac{\partial^{2}V_{eff}}{\partial\phi^{2}}(0,r)\phi,\\
&&B(r)=-\frac{1}{2}\log\left(1-\frac{M}{4\pi r}+\frac{g^{2}}{2r^{2}}\right),\quad \frac{\partial^{2}V_{eff}}{\partial\phi^{2}}(0,r)=m^{2}-\frac{g^{2}\ell^{2}}{r^{4}},\\
&&\phi\propto 1+\frac{2(g^{2}\ell^{2}-m^{2})}{g^{2}-2}(r-1)\quad\mbox{for}\quad r\to1,\\
&&\phi\propto\frac{\rme^{-mr}}{r}\quad\mbox{for}\quad r\to\infty
\end{eqnarray*}
where $M$ is the mass of the black hole. We refer to \cite{gup} for the determination of $M$ by fitting $\rme^{-2B}$ to the Reissner-Nordstr\"{o}m form for large $r$. Taking into account that the requirement $r_{H}=1$ is equivalent to require that $M=4\pi+2g^{2}$ the ODE for $\phi$ can be brought into the form
\[
\phi^{''}+p(r)\phi^{'}+q(r)\phi=0
\]
with
\[
\fl
p(r)=\frac{p_{H}}{r-1}+\frac{p_{-}}{r-r_{-}},\quad p_{H}=\frac{2}{1-r_{-}},\quad p_{-}=-r_{-}p_{H},\quad r_{-}=\frac{g^{2}}{2\pi},
\]
\[
\fl
q(r)=-m^{2}+\frac{q_{00}}{r^{2}}+\frac{q_{0}}{r}+\frac{q_{H}}{r-1}+\frac{q_{-}}{r-r_{-}},
\]
\[
\fl
q_{00}=2\pi\ell^{2},\quad q_{0}=\left(1+\frac{1}{r_{-}}\right)q_{00},\quad q_{H}=\frac{g^{2}\ell^{2}-m^{2}}{1-r_{-}},\quad q_{-}=\left(m^{2}r_{-}-\frac{q_{00}}{r_{-}}\right)\frac{1}{1-r_{-}}.
\]
Finally, by means of the transformation
\[
\phi(r)=r^{\widetilde{\alpha}}\rme^{-mr}\psi(r)
\]
with $\widetilde{\alpha}$ such that $\widetilde{\alpha}^{2}-\widetilde{\alpha}+q_{00}=0$ it can be checked that $\psi$ satisfies the GHE (\ref{GHE}) with 
\begin{eqnarray*}
&&\mu_{0}=1-2\widetilde{\alpha},\quad \mu_{1}=1-p_{H},\quad \mu_{2}=1-p_{-},\quad\alpha=-2m,\\
&&\Lambda_{0}=q_{0},\quad\Lambda_{1}=q_{H},\quad\Lambda_{2}=q_{-}.
\end{eqnarray*}
\section{Conclusions}
In this paper we have related the radial Dirac equation for a massive fermion in the Kerr-Newman geometry to a GHE. In addition, we found that the GHE describes also static perturbations for the non-extremal Reissner-Nordstr\"{o}m solution. We reserve the study of other examples where the GHE appears for future investigations.
\ack
The research of D.B. was supported by the EU grant HPRN-CT-2002-00277. D.B. thanks Prof. Steven Gubser, Department of Physics, Princeton University for useful explanations concerning the Lagrangian introduced in Section~\ref{sec:3}. M.W. was supported by the DFG (German Research Fund), grant TR 368/6-1.

\end{document}